# Investigation of asymmetrically pitching airfoil at high reduced frequency

Mitesh Thakor, Gaurav Kumar, Debopam Das, and Ashoke De[*]

*Department of Aerospace Engineering, Indian Institute of Technology Kanpur, Kanpur 208016, India*

*corresponding author email address: ashoke@iitk.ac.in

**Abstract**

The expanding application in Micro-Air Vehicles has encouraged many researchers to understand the unsteady flow around a flapping foil at a low Reynolds number. We numerically investigate an incompressible unsteady flow around a two-dimensional pitching airfoil (SD7003) at high reduced frequency ($k \geq 3$) in the laminar regime. This study interrogates the effect of different unsteady parameters, namely amplitude (A), reduced frequency ($k$), Reynolds number (Re), and asymmetry parameter (S) for pitching motion on the force coefficients. The inviscid theoretical model is utilized to calculate the lift coefficient for sinusoidal motion in the viscous regime, and a comparison is made with the numerical results. The theoretical analysis identifies the influence of the non-circulatory lift over circulatory lift at a high reduced frequency. Further, the results indicate that the reduced frequency ($k$) and asymmetry parameter (S) have a significant impact on the instantaneous and time-averaged force coefficients as well as on the vortex structure in the wake. Finally, the Fast Fourier Transformation (FFT) analysis is carried out over a simulated case with fixed amplitude and Reynolds number for distinct $k$ and S values. The findings confirm that the dominant frequency in the flow ($k^*$) has a direct correlation to the airfoil pitching frequency ($k$).

**Keywords:** Low Reynolds number; Pitching airfoil; Asymmetric sinusoidal motion; Unsteady aerodynamics

1. INTRODUCTION

For the last couple of decades, the progress in the development of Micro-Air Vehicles (MAVs) and Unmanned Arial Vehicles (UAVs) has prompted scientists and engineers to understand the flow dynamics of a flapping wing at low Reynolds number. The aerodynamics characteristic of the flapping foil depends on the various unsteady parameters, namely amplitude, flapping frequency, type of pitching motion, Reynolds number. A few researchers extensively reviewed the influence of the unsteady parameters with low Reynolds numbers[1, 2]. In addition to the unsteady parameters, McCroskey et al.[3, 4] experimentally showed the impact of the vortex structure around the airfoil (leading edge and trailing edge vortex) on the force coefficients. The reduced frequency ($k = \omega C/2U_\infty$; $\omega$ = angular pitching frequency, C = Chord length of the airfoil, $U_\infty$ = incoming flow velocity) is one of the significant parameters which governs the flow physics of the flapping foil. Many researchers have examined the low reduced frequency flapping airfoil at low Reynolds number (LRN). For example, Mehta[5] numerically showed that the lift coefficient contribution is due to the pressure at low Reynolds number (Re = 5000 and 10000) at low reduced frequency ($k = 0.5$ and 0.25). Whereas, Akbari and Price[6] reported delayed flow separation with the increase in reduced frequency (in turn affecting the force coefficients) while conducting the two-dimensional (2D) study of pitching airfoil (modified NACA0012) using the vortex method for LRN. In an experimental study at a relatively higher reduced frequency ($k = 1$), Ohmi et al.[7] qualitatively observed that the vortex wake pattern changes and depends on different unsteady parameters while investigating the flow field over NACA0012 as well as on an elliptical airfoil in the range of Re = 1500 to 10000.

The effect of unsteady parameters, Reynolds number and motion type, have been investigated by different researchers on the force coefficients. The significant studies are Kim and Chang[8], Amiralaei et al.[9], Schouveiler et al.[10], Cleaver et al.[11]. The most notable one is the study of Kim and Chang[8], where they investigated the effect of Reynolds number on sinusoidally pitching NACA0012 airfoil at a low reduced frequency ($k = 0.1$). They observed the non-linear behavior of the lift coefficient and pressure coefficient, which was primarily due to the nonlinear response of the unsteady boundary layer at high Re. For different Reynolds numbers, Amiralaei et al.[9] numerically and analytically studied 2D flow around the pitching airfoil in the LRN regime. Their theoretical results (Theodorsen's method) differed from the numerical results for high viscous flow (at low Re) with low reduced frequency ($k < 0.1$). On the other hand, the experimental investigation of Schouveiler et al.[10] reported the combined pitching and plunging motion for a fixed Re = 40000. Here, the authors concluded that a cosine motion generated the most efficient thrust, and the propulsive behavior of the flapping was due to jet type wake in the downstream. In addition to the plunging motion, for small amplitude and high Strouhal number, Cleaver et al.[11] experimentally indicated that the shear layer rolls into multiple coherent vortices for high pitching frequency at Re = 10000.

Moving forward to the high-frequency oscillations, both experimental as well as computational studies show that the flow physics completely changes at high reduced frequency pitching foil as compared to low frequency. The earlier work by Garrick[12] reported that the reverse von Karman vortex street is observed for a high $k$; as a result, the velocity deficit (wake-type, drag dominated) profile in the wake transforms to a velocity excess (jet-type, thrust generating) profile. Later, for the high reduced frequency ($k > 2$), Koochesfahani[13] experimentally calculated the drag coefficient based on the mean velocity profile for NACA0012 at Re 12000. The author concluded that for a particular combination of parameters (Reynolds number, reduced frequency), the velocity deficit profile changed to velocity excess as $k$ increased. For non-sinusoidal motion, their reported results (qualitative results only) showed the formation of a complex vortex structure while comparing the same with a case with the sinusoidal motion for the same set of parameters. The numerical investigation by Ramamurti and Sandberg[14] furthermore showed that the critical parameter for the thrust generation is not $k$ but the Strouhal number based on trailing edge displacement. Bohl and Koochesfahani[15] experimentally examined the high reduced frequency ($k > 2$) pitching airfoil for LRN (Re 12000), and reported that the spatial arrangement of the vortex array indeed depended on the reduced frequency.

In addition to the basic set of parameters (Reynolds number, reduced frequency and motion type), the other essential parameters (which influence the force coefficient) are pitching wave motion and the shape of the airfoil. In a numerical investigation using panel code, La Mantia and Dabnichki[16] concluded that the thrust coefficient reduces for a fixed-wing mass with the increase in airfoil thickness. McGowan et al.[17] carried out an experimental study over SD7003 (low Re airfoil) airfoil at a high reduced frequency ($k$

= 3.93) for pure pitching motion at two different Reynolds number 10000 and 40000, and concluded a negligible Re effect on the flow structure at high *k*. Later, Ol et al.[18] observed a discrepancy between experiment and theory for separated flow over the SD7003 for a shallow and deep stall at Re 60000. Lua et al.[19], while studying the elliptical airfoil in a forward flight for Re = 5000 with a combined airfoil motion, reported the presence of multiple peaks in the thrust generation due to the high rotation rate of the airfoil. The few handful works of literature are available which investigate the influence of a pure non-sinusoidal pitching motion of the airfoil. However, there are some studies where authors reported the impact of the non-sinusoidal motion on the flow physics and the force coefficients with a pitch-plunge combined or a pure-plunge airfoil motion. One such study includes the work of Ashraf et al.[20] for the plunge-pitch combined motion with LRN. The authors indicated that the propulsion efficiency significantly increased for non-sinusoidal motion as compared to the sinusoidal motion. Sarkar and Venkatraman[21] reported similar observations for the asymmetric sinusoidal motion for pure plunging oscillation and found that the thrust was minimum for typical sinusoidal plunging motion as compared to asymmetric sinusoidal motion for the same *k*. For NACA0012 airfoil undergoing asymmetric pitching motion at high Reynolds number (1.35 × $10^5$), Lu et al.[22] numerically showed that the asymmetry parameter (S) considerably affects the force coefficients for the same reduced frequency. Recently, the computational study for non-sinusoidal pitching motion at LRN by Das et al.[23] arrived at the similar conclusion that the thrust force improves for a non-sinusoidal motion.

Apart from the experimental or computational, one can alternatively model the flapping foil behavior using the theoretical model. However, any theoretical model has its limitations and assumptions. The wide variety of theoretical models were developed and modified over time for a dynamic stall to capture a reasonably accurate aerodynamic characteristic of the flapping foil. Initially, the analytical theory for unsteady aerodynamics was developed for aircraft flutter problems (Theodorsen[24] and von Kármán[25, 26]); however, it has acquired an immense application in the bio-fluid mechanics and the insect flapping motion. Petot[27] developed a theoretical model at the Office National D'Etudes et de Recherches Aerospatiales (ONERA(EDLin)) for attached and separated inviscid flow for pitching airfoil motion and later extended it to the plunging and pulsating incoming flow in the extended ONERA(EDLin) model[28]. Later, Laxman and Venkatesan[29] used the extended ONERA(EDLin) model and showed that the second-order approximation (Venkatesan and Friedmann[30]) of the Theordorson's lift deficient function gave a better result compared with the experimental data at high Reynolds number.

From the brief literature review, it seems that the comprehensive studies have been carried out for the low reduced frequency airfoil for both high and low Reynolds numbers. Besides, very few studies reveal the effect of the high reduced frequency with LRN. However, at high-frequency pitching airfoil, the past studies have given the primary focus to the thrust coefficient, while the lift coefficient remains unexamined. Also, the effect of an asymmetric parameter in the pitching wave motion for the low Reynolds number flow is not addressed thoroughly. Moreover, the theoretical model was adopted mostly for the high Reynolds number flow, and the capability of an inviscid model (theoretical model) over a viscous regime remains unanswered for high-frequency pitching airfoil. Therefore, the present study aims to address the said concerns, while utilizing a similar theoretical model to emphasize the applicability of the inviscid theory model to calculate the lift coefficient of pitching airfoil for viscous flow at low Reynolds number. In this work, we present a comprehensive study over low Reynolds number airfoil (SD7003) for asymmetric sinusoidal motion in comparison to the normal sinusoidal motion for different unsteady parameters [amplitude(A), reduced frequency($k$), and asymmetry parameter(S)] for two Reynolds number, i.e., 3000 and 10000.

The article is organized as follows: the numerical and theoretical models are discussed in section 2. Section 3 presents the computational details and validation studies. The final results are discussed in section 4, followed by the concluding remarks in section 5.

## 2. NUMERICAL AND THEORETICAL MODEL

### 2.1 Numerical Method

The governing equations for unsteady, two-dimensional, incompressible, laminar flow around SD7003 are listed below in the non-dimensional form

$$\frac{\partial u_i}{\partial x_i} = 0 \tag{1}$$

$$\frac{\partial u_i}{\partial t} + \frac{\partial u_i u_j}{\partial x_j} = -\frac{\partial p}{\partial x_i} + \frac{1}{Re}\frac{\partial^2 u_i}{\partial x_i \partial x_j} \tag{2}$$

The above equations are solved using the finite volume method in the OpenFOAM framework [OpenFOAM user guide[31]]. The convective flux discretization deploys second-order TVD (Total Variation Diminishing) scheme, while the viscous flux discretization involves second-order central scheme, and the temporal term deploys the first-order implicit Euler scheme with sufficiently small time steps to maintain stability and reduce numerical diffusion. The pressure-velocity coupling is obtained by using the PIMPLE algorithm. Parallel processing is achieved by the message passing interface (MPI) technique. All the computations are carried out while keeping the Courant number below 0.5.

### 2.2 Theoretical Formulation (inviscid assumption)

The analytical model, extended ONERA(EDLin)[28] dynamic stall model, is invoked to calculate the unsteady aerodynamic lift over the sinusoidally pitching airfoil. Venkatesan and Friedmannn[30] modified model for higher-order (second-order) approximation for Theodorsen lift deficiency function. The stall model assumes that forces and moment act at the quarter chord point. The unsteady lift acting on the airfoil is given by

$$L = \frac{1}{2}\rho\tilde{S}[sb\dot{W}_0 + \tilde{k}b\dot{W}_1 + U_\infty\Gamma_1 + U_\infty\Gamma_2] \tag{3}$$

$$\dot{\Gamma}_1 + \lambda\left(\frac{U_\infty}{b}\right)\Gamma_1 = \lambda\left(\frac{U_\infty}{b}\right)\frac{\partial C_{zL}}{\partial\alpha}W_0 + \lambda\sigma\left(\frac{U_\infty}{b}\right)W_1 + \left(\tilde{\alpha}\frac{\partial C_{zL}}{\partial\alpha} + d\right)\dot{W}_0 + \tilde{\alpha}\sigma\dot{W}_1 \tag{4}$$

$$\ddot{\Gamma}_2 + a\left(\frac{U_\infty}{b}\right)\dot{\Gamma}_2 + r\left(\frac{U_\infty}{b}\right)^2\Gamma_2 = -\left[r\left(\frac{U_\infty}{b}\right)^2 U_\infty \Delta C_z|_{W_0/V} + E\left(\frac{U_\infty}{b}\right)\dot{W}_0\right] \tag{5}$$

The various coefficients of the equations (3, 4, and 5) can be found from ref. 29 and the coefficient for $C_{zl}$ is taken from the static data presented by Anyoji et al.[32] for SD7003.

In the present study, the maximum amplitude ($A_{max}$) is 6°, which is an unstalled condition. Therefore, in the equation (3), $\Gamma_2$ is dropped for the unstalled condition and $\Delta C_z$ is taken as zero[29]. The stall model is furthermore applied only to the harmonic motion, $\alpha = \alpha_0 + A\, e^{i\omega t}$, of the airfoil. With these considerations, the final form of equation (3) is listed below. The Lift force is expressed in two terms namely circulatory lift ($L_C$) and non-circulatory lift ($L_{NC}$) which are given by

$$L = L_{NC} + L_C \tag{6}$$

$$L_{NC} = \frac{1}{2}\rho\tilde{S}\left[\pi b U_\infty \dot{\alpha} + \frac{\pi}{2}b^2\ddot{\alpha}\right] \tag{7}$$

$$L_C = \frac{1}{2}\rho\tilde{S}U_\infty(2\pi)C(k)\left[U_\infty\alpha_0 + U_\infty A e^{i\omega t} + b\dot{\alpha}\right] \tag{8}$$

Where $C(k) \cong \dfrac{A_1\left(\frac{i\omega b}{U_\infty}\right)^2 + A_2\left(\frac{i\omega b}{U_\infty}\right) + A_3}{\left(\frac{i\omega b}{U_\infty}\right)^2 + B_2\left(\frac{i\omega b}{U_\infty}\right) + B_3}$, second order approximation[30] to exact Theodorsen lift deficiency function ($C(k)$). Here $A_1 = 0.50$, $A_2 = 0.393$, $A_3 = 0.0439425$, $B_2 = 0.5515$ and $B_3 = 0.0439075$.

The total lift coefficient is given by equation (9), which is a summation of the non-circulatory lift coefficient ($C_{L,NC}$) and circulatory lift coefficient ($C_{L,C}$).

$$C_L = C_{L,NC} + C_{L,C} \tag{9}$$

$$C_{L,NC} = \frac{L_{NC}}{\frac{1}{2}\rho\tilde{S}U_\infty^2} \text{ and } C_{L,C} = \frac{L_C}{\frac{1}{2}\rho\tilde{S}U_\infty^2}$$

**2.3 Kinematics of the airfoil motion**

The extensive study exists for the typical sinusoidal motion of the airfoil with higher reduced frequency, numerically as well as experimentally. The typical sinusoidal motion of the airfoil is given by

$$\alpha(t) = \alpha_0 + A\cos(2\pi f t) \tag{10}$$

Here $\alpha_0$ is the mean angle of attack (AoA), and A is the maximum angle of attack. In the present study, the sinusoidal motion is compared with the asymmetric sinusoidal motion and the mean $\alpha_0$ is taken as zero. The asymmetry parameter S is introduced into the equation of motion to achieve asymmetric cosine wave, as shown in Fig. 1(a). The parameter S is defined as the fraction of the time spent from one trough to the next crest with respect to the time period of one cycle. Hence, for S = 0.5, relation 11 gives the normal sinusoidal motion. For the smaller values of S, i.e., S = 0.3 and 0.4, the pitching upcycle takes less time as compared to pitching down cycle implies that the pitch-up cycle becomes faster, and the pitch-down cycle becomes slower. The asymmetric sinusoidal motion is represented by

$$\alpha(t) = A \sin\left(\frac{2\pi}{2ST} t - \frac{\pi}{2}\right), \quad 0 \leq t \leq ST$$

$$\alpha(t) = A \sin\left(\frac{2\pi}{2(1-S)T} (t - T) - \frac{\pi}{2}\right), \quad ST \leq t \leq T$$

(11)

Where T is the time period for one cycle. Initially, the airfoil is at the minimum amplitude ($A_{min}$), and the pitching axis for all cases is at quarter chord from the leading edge. The effect of asymmetry parameter S on the airfoil's angular acceleration ($\ddot{\alpha}$) is shown in Fig. 1(b) for A = 2°. The normalized time ($\tau$) is given by $\tau = t / T$. All the cases considered for the present study are listed in Table I.

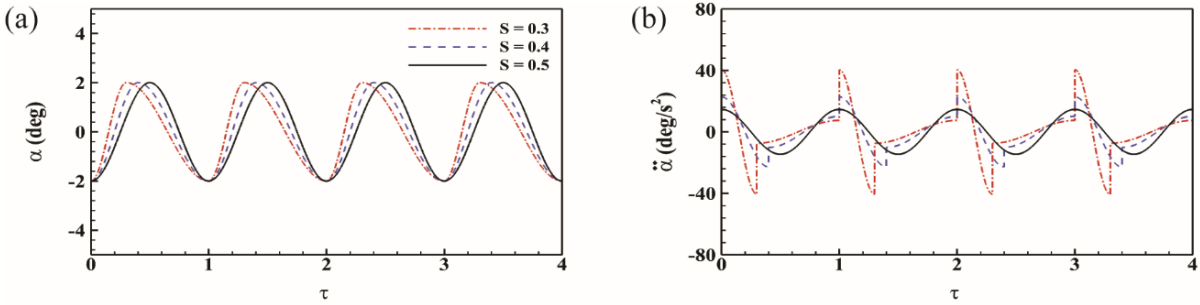

FIG. 1. (a) The pitching motion for the different asymmetry parameter S for A = 2°. (b) Angular acceleration of the pitching airfoil corresponding to its pitching motion.

TABLE I. Simulation parameters considered in the present study.

| Reynolds Number | Amplitude | Reduced Frequency | Asymmetry parameter |
|---|---|---|---|
| (Re) | (A) | (k) | (S) |
| 3000 and 10000 | 2° | 3, 6, 9 | 0.3, 0.4, 0.5 |
|  | 4° | 3, 6, 9 | 0.3, 0.4, 0.5 |
|  | 6° | 3, 6, 9 | 0.3, 0.4, 0.5 |

## 3. BOUNDARY CONDITIONS AND VALIDATION STUDY

Figure 2 depicts the simulation domain for the present study with boundary conditions. A high-quality structured, multiblock grid is prepared by using ANSYS ICEM CFD® with two mesh regions, one is the inner rotating grid, and the other is the outer stationary grid. The sliding interference is accommodated by cyclic-AMI boundary condition, which gives freedom to provide unequal grid-points over the interface of the inner and outer circle grid. Thus, a coarse grid can be used at far-field to reduce the computational time and resources. A constant velocity inlet boundary condition is provided at the inlet and the fixed pressure for outgoing flow at the outlet. The no-slip condition is applied to the airfoil, and the diameter of the inner rotating grid is 10C (C is the chord length of the airfoil). The upper and lower boundaries are set to be far-fields (free stream), which are located at 15C from the airfoil chord line. The inlet and outlet boundaries are at 15C and 20C, respectively, from the quarter chord point, which ensures a negligible boundary effect on the airfoil boundary layer.

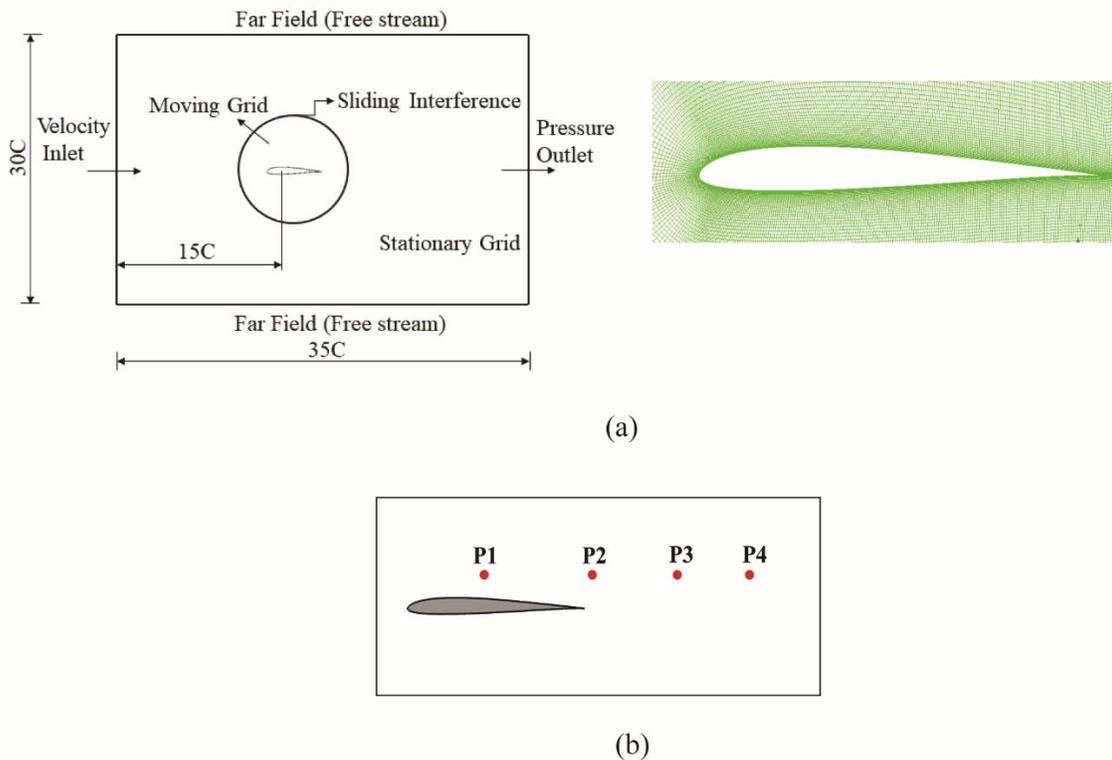

FIG. 2. (a) A schematic of the computation domain with the boundary conditions and the close-up view of the structural grid around the SD7003 airfoil. (b) A schematic of the probe's locations in the domain.

We validate the solver and the grid for two different cases over a range of reduced frequencies. The numerical simulations are carried out on three separate grids, coarse G1 (0.051M cells), medium G2 (0.13M cells), and fine G3 (0.158M cells) grids for a high reduced frequency. The $y^+$ value is maintained below 1 for all the grids. For Re = 12000, the flow conditions and kinematics of the pitching airfoil (A = 2°) are consistent with the experimental study by Koochesfahani[13] for NACA0012. The simulations are carried out only for four cycles to reduce the computational effort as the force coefficients are not changing after the second pitching cycle. The time-averaged mean drag coefficient (calculated based on the fourth cycle) is shown for three different reduced frequencies with the percentage change over coarse G1 to medium G2 and medium G2 to fine G3 grid (Table II). As referred from Table II, the changes in the mean drag coefficient over medium to the fine grid are negligible for all frequencies. Hence, the medium grid is found to be sufficient for the detailed numerical calculations. The present results (medium grid G2) are furthermore compared with the experimental result of Koochesfahani[13] and the previous numerical results by Ramamurti and Sandberg[14] (Table III). Table III shows that for the present study, the mean drag coefficient ($\overline{C_D}$) is in a close agreement with the experimental results for various reduced frequencies and significantly improved from previous numerical results.

For further validation, the qualitative results for low pitching reduced frequency ($k$ = 0.1) and low Reynolds number (Re 3000) on the medium grid G2 are compared with the flow visualization experimental results (Ohmi et al.[7]). Figure 3 illustrates the snapshots of the streamlines of the unsteady vortex wake around the airfoil along with the experimental collection over the same time instance. The pitching axis is located at the semi-chord point, and the mean incidence and angular amplitude of the pitching, both are 15°. The streamlines around the airfoil for all the time instances are in good agreement with the experimental visualization by Ohmi et al.[7]. The leading edge and trailing edge vortices are captured very well, as shown in Fig. 3. The comparison indicates that for the low Reynolds number and high angle of attack pitching, the medium grid G2 is sufficient to capture the flow pattern.

TABLE II. The time-averaged mean drag coefficient ($\overline{C_D}$) compression over the different grids; coarse G1 (0.051M cells), medium G2 (0.13M cells), and fine G3 (0.158M cells) with percentage change over the grids for different reduced frequency.

| | Mean Drag Coefficient ($\overline{C_D}$) | | | Percentage change | |
|---|---|---|---|---|---|
| $k$ | Coarse G1 (0.051M) | Medium G2 (0.13M) | Fine G3 (0.158M) | G1 to G2 | G2 to G3 |
| 1.8 | 0.0326 | **0.0326** | 0.0326 | 0 | 0 |
| 3.2 | 0.0292 | **0.0298** | 0.0299 | 2.054 | 0.33 |
| 5.1 | 0.0224 | **0.0227** | 0.0226 | 1.34 | -0.44 |

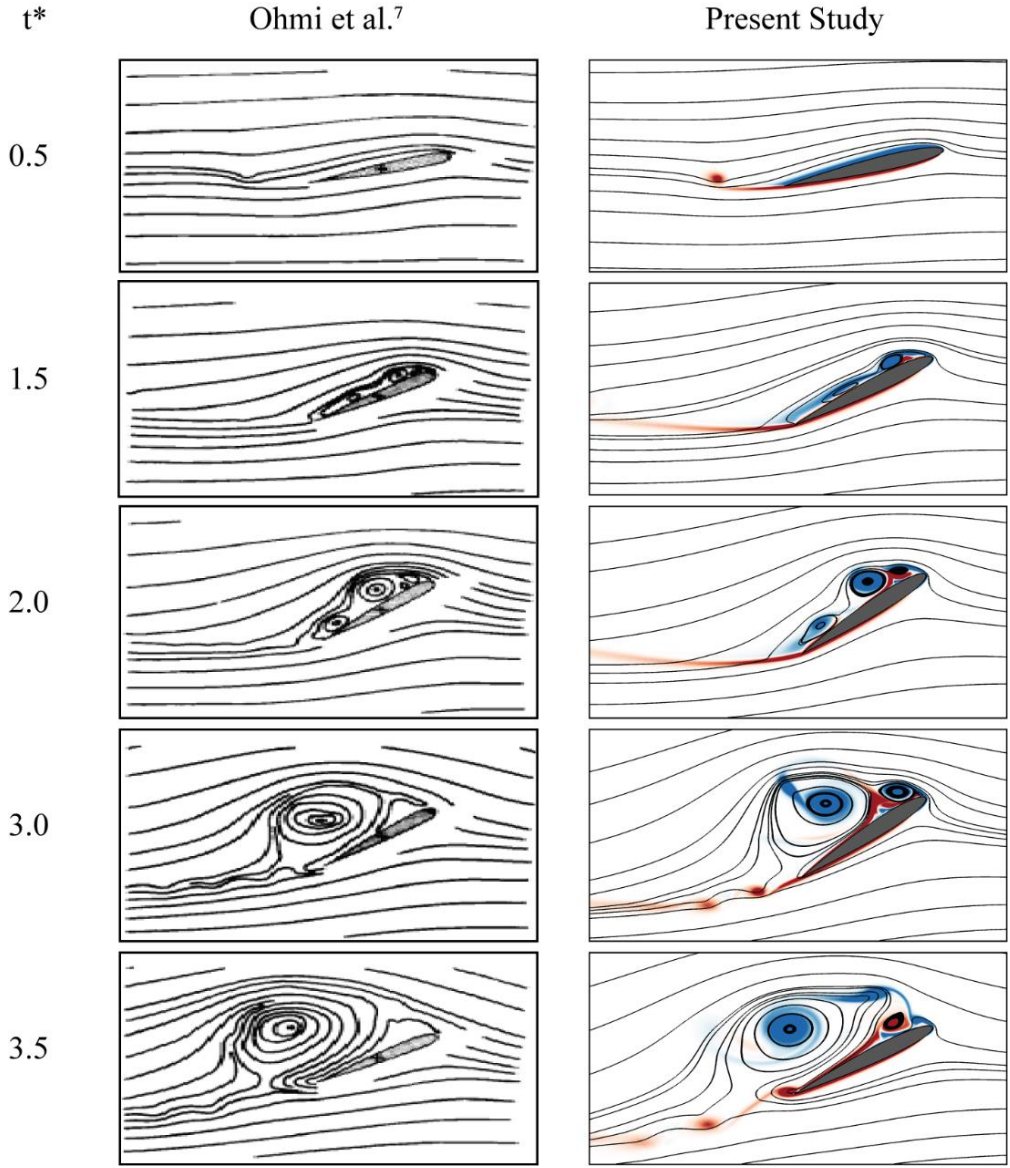

FIG. 3. The comparison of experimental (Ohmi et al.[7]: Reproduced with permission from Journal of Fluid Mechanics, 225 (1991): 607-630") instantaneous unsteady streamlines patterns around the NACA0012 at different time intervals with the present study.

TABLE III. The time-averaged mean drag coefficient ($\overline{C_D}$) comparison with the experimental results by Koochesfahani[13] and the numerical results by Ramamurti and Sandberg[14] for the present study on the medium grid G2 (0.13M) for three pitching frequencies.

| | Mean Drag Coefficient ($\overline{C_D}$) | | |
|---|---|---|---|
| $k$ | Koochesfahani[13] | Ramamurti and Sandberg[14] | Present Study (medium grid G2) |
| 1.8 | 0.0335 | 0.0397 | 0.0326 |
| 3.2 | 0.0306 | 0.0385 | 0.0298 |
| 5.1 | 0.015 | 0.0331 | 0.0227 |

## 4. RESULTS AND DISCUSSION

This section elucidates the simulation results for pitching SD7003 airfoil at a high reduced frequency and low Reynolds number. The different unsteady parameters are analyzed and aim to interrogate the influence of an asymmetry sinusoidal motion of the airfoil. The results of the present study are given into two major parts: first part illustrates the comparison between the numerical and the theoretical results (section 4.1), while section 4.2 presents the effect of different Reynolds number. The instantaneous and time-averaged force coefficients are investigated in sections 4.3 & 4.4, respectively, for various unsteady parameters. The results in the second part examine the vorticity formation for different S values (section 4.5), followed by the supported data of Fast Fourier Transformation (section 4.6) at the locations of the different probes (Fig. 2).

### 4.1 Theoretical result (Effect of pitching amplitude)

For the steady-state response, equation (8) is converted to the state-space form, and the fourth-order Runge-Kutta time integration scheme has been used[29]. The initial aerodynamic states are assumed to be zero. The brackets < > shows the force coefficients over the fourth cycle. The theoretical result is obtained for the symmetrically pitching airfoil.

The simulations are conducted for all the combinations of parameters, which are mentioned in Table I. Only the most significant results are shown and discussed here. The total theoretical lift coefficient is compared with the numerical result for the parameters, A = 2°, and 6°, Re = 3000 and 10000, and pitching frequency $k$ = 9 (Fig. 4). The comparison of the lift coefficient components (refer equation 9), non-circulatory and circulatory lift coefficient, entails that the non-circulatory lift coefficient is preeminent over the circulatory lift coefficient at the higher reduced frequency pitching motion. The reason behind such domination is that the non-circulatory lift is directly proportional to the angular acceleration of the airfoil (refer equation 7), and the angular acceleration varies quadratically with the angular frequency ($\sim k^2$), implying that as reduced frequency ($k$) increases, the acceleration of the airfoil increases tremendously. Hence the contribution of the non-circulatory lift is much higher than the circulatory lift. At the starting of the cycle (at $\alpha = A_{min}$), the acceleration of the airfoil is maximum (Fig. 1(b)), so at that instant, the $C_{L,NC}$ will be maximum. As airfoil is going from $A_{min}$ to $A_{max}$, the acceleration decreases; the effect can be seen in $C_{L,NC}$ which also decreases. After reaching the max AoA, the airfoil starts pitching down from where acceleration will increase so that $C_{L,NC}$ increases. On the other hand, the circulatory lift ($C_{L,C}$) has a nominal value, and as typically observed for the low reduced frequency ($k < 1$) pitching airfoil, it increases with increasing AoA. The non-circulatory lift coefficient even dominates the shape of the total lift coefficient hysteresis curve. The results are similar for the other parameters combination for both Re 3000 and 10000. The theoretical total lift coefficients are in good agreement with the numerical result for low amplitude A = 2° and high reduced frequency $k = 9$ for both Re 3000 and 10000. However, at the high amplitude (i.e., A = 6°), the theoretical method underpredicts the total lift coefficient compared to the numerical study for both the Reynolds numbers 3000 and 10000 (Fig. 4). The cause of a difference is that the analytical model was developed considering a small amplitude perturbation. At higher reduced frequency, force coefficients are mostly dominated by non-circulatory lift, enhancing the role of high amplitude perturbation. This puts severe restrictions on the use of the theoretical model for high amplitude and high-frequency pitching motion calculations.

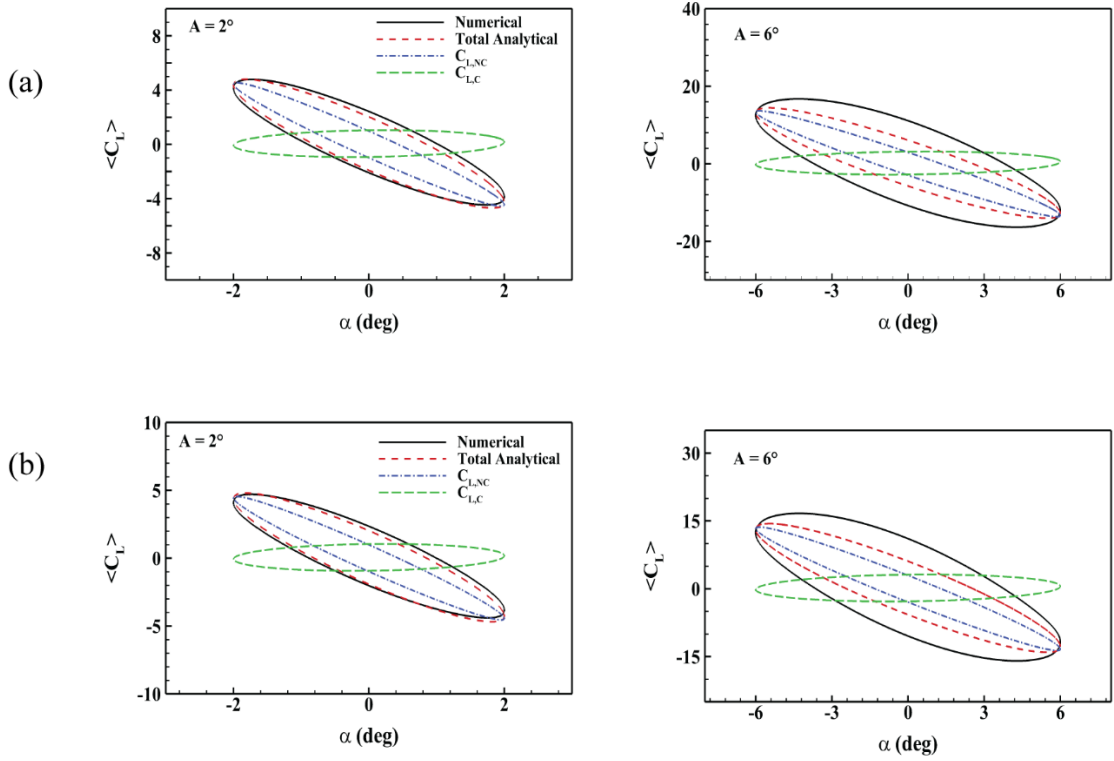

FIG. 4. The theoretical total lift coefficient with its components, non-circulatory lift coefficient ($C_{L,NC}$) and circulatory lift coefficient ($C_{L,C}$) are compared with the numerical result for A = 2° and 6°, $k$ = 9. (a) Re = 10000 and (b) Re = 3000. The brackets < > shows the force coefficients over the fourth cycle.

TABLE IV. Percentage difference for the maximum and minimum lift coefficients ($C_{L,max}$ and $C_{L,min}$) over the Re 3000 and 10000 for the parameter A = 4°, S = 0.5, and different $k$ values.

|  | $k = 3$ | | $k = 9$ | |
| --- | --- | --- | --- | --- |
|  | $C_{L,max}$ | $C_{L,min}$ | $C_{L,max}$ | $C_{L,min}$ |
| **Re 3000** | 1.3209 | -1.0469 | 10.2423 | -9.7256 |
| **Re 10000** | 1.4407 | -1.1272 | 10.3772 | -9.8796 |
| **% difference** | 8.31 | 7.12 | 1.30 | 1.55 |

TABLE V. Percentage difference for the maximum and minimum lift coefficients ($C_{L,max}$ and $C_{L,min}$) over the Re 3000 and 10000 for the parameter A = 4°, S = 0.3, and $k$ = 9.

|  | $k = 9$ | |
| --- | --- | --- |
|  | $C_{L,max}$ | $C_{L,min}$ |
| **Re 3000** | 25.9101 | -24.6775 |
| **Re 10000** | 26.1256 | -25.1318 |
| **% difference** | 0.82 | 1.80 |

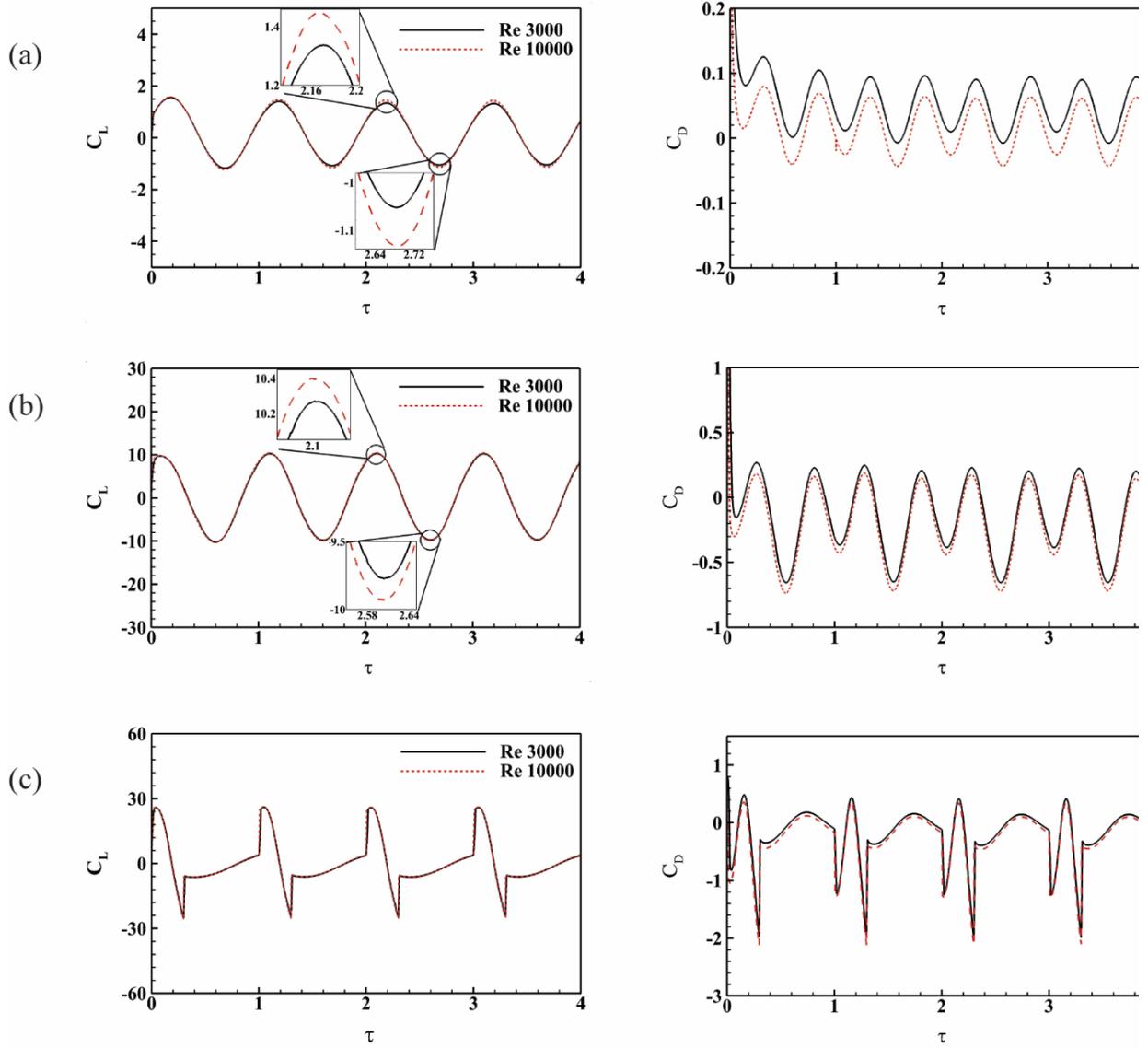

FIG. 5. The instantaneous force coefficients ($C_L$ and $C_D$) comparison for Reynolds number 3000 and 10000 at A = 4°. (a) $k = 3$, S = 0.5; (b) $k = 9$, S = 0.5 and (c) $k = 9$, S = 0.3.

### 4.2 Reynolds number effect

Here the analysis is extended for two different Reynolds numbers, i.e., 3000 and 10000, to address the impact of Reynolds number on the force coefficients at a higher reduced frequency ($k \geq 3$) for the pitching airfoil. Figure 5 reports the instantaneous force coefficients with normalized time ($\tau = t / T$) for the parameters $A = 4°$, $k = 3$ and 9 and two distinct S values 0.5 and 0.3. The time history of $C_L$ is approximately identical for both Reynolds number 3000 and 10000 for all S values. The maximum and minimum lift coefficients, $C_{L,max}$, and $C_{L,min}$, respectively, are magnified in Figs. 5(a) and 5(b) for symmetric motion (S = 0.5) as well as for the quantitative comparison, the percentage difference for maximum and minimum lift coefficients over Re are listed in Table IV. The $C_{L,max}$, and $C_{L,min}$ have a comparatively moderate difference over Re 3000 and 10000 at the lower pitching reduced frequency ($k = 3$), but with increasing the $k$ value, the percentage difference between the peaks of the lift coefficients diminishes (Table IV). These observations indicate that at a higher reduced frequency ($k = 9$), the maximum and minimum lift generation is mostly dependent on the inertial forces due to the acceleration of the airfoil, and Reynolds number has a negligible effect on the peaks of the $C_L$. This is also in agreement with the discussion in the previous section, the lift generation for high pitching rate airfoil is mainly due to the non-circulatory component where the viscous wake effect (circulatory lift) on the lift force is negligible. However, the Reynolds number effect is more significant on the drag coefficient ($C_D$). At low reduced frequency ($k = 3$), the drag coefficient is higher for Re = 3000 as compared to Re = 10000, due to higher viscous effect at low Reynolds number. Increasing reduced frequency only makes inertial forces stronger, resulting in viscous effect becoming negligible compared to inertial forces at the $k = 9$.

Therefore, $C_D$ varies approximately the same for both Re, as shown in Fig. 5(b). In Fig. 5(c), this same feature is observed for the force coefficients over the Reynolds numbers at $k = 9$ despite the asymmetric pitching motion with $S = 0.3$ (see Table V for quantitative comparison). Hence, for further analysis, results are discussed only for Re = 10000, and the same features are obtained for different Reynolds numbers.

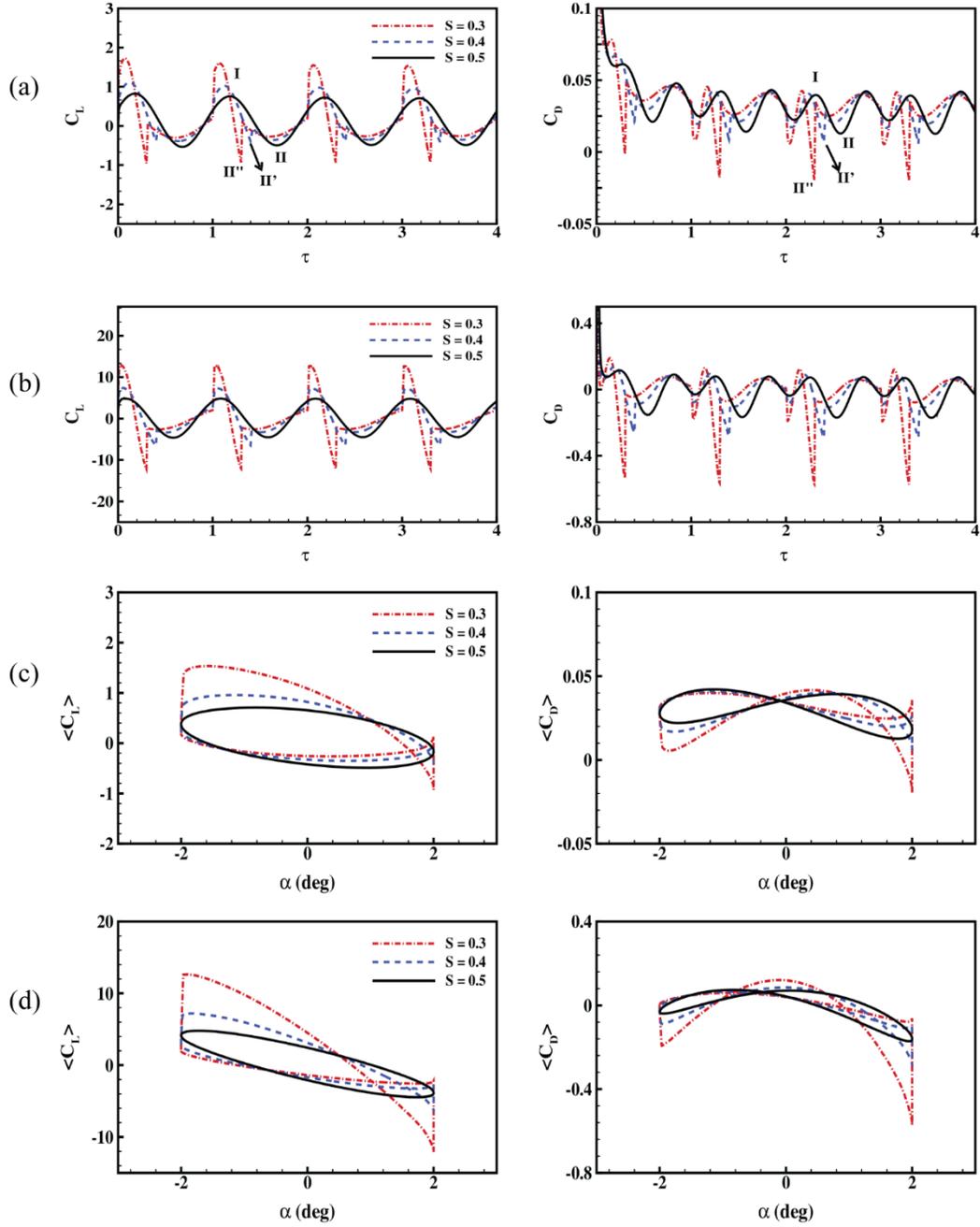

FIG. 6. The force coefficient variation for different asymmetry parameters (S) for the same reduced frequency (*k*) and amplitude (A) with normalized time (τ) and Angle of Attack (α) with A = 2° and Re 10000. (a) and (c) *k* = 3; (b) and (d) *k* = 9. The brackets < > shows the force coefficients over the fourth cycle.

### 4.3 Force coefficients

Figure 6 reports the time history of the instantaneous force coefficients ($C_L$ and $C_D$) for A = 2° with *k* = 3 and 9 at Re 10000. After initial fluctuations (The airfoil starts from rest at minimum AoA so the sudden acceleration of the airfoil would cause the initial oscillation.) during the first cycle, the flow and force coefficients have become periodic. As a result, to reduce the computation time and cost, the simulations are carried only first four cycles, and the force coefficients ($C_L$, $C_D$) are plotted against the angle of attack ($\alpha$) for the fourth cycle which is indicated by brackets $<>$.

The peaks of the force coefficients increase as the airfoil goes from symmetric ($S = 0.5$) to asymmetric ($S = 0.3$) sinusoidal motion (Figs. 6(a) & 6(b)). The maximum lift coefficient ($C_{L,max}$) (indicated by I in Fig. 6(a)) occurs immediately after the pitch-up motion for all S values and the phase differences among the S values for $C_{L,max}$ is nominal. On the contrary, the occurrence and the magnitudes of the minimum lift and drag coefficient ($C_{L,min}$ and $C_{D,min}$) depend on the S parameter (mentioned by II, II', II'' for $S = 0.5$, 0.4 and 0.3, respectively in Fig. 6(a)). The increment in the peaks of the force coefficients for an asymmetric motion has a massive impact on the mean force coefficients that will be discussed in the next section. For the asymmetry parameter $S = 0.3$, the time taken for the pitch-up cycle is less as compared to $S = 0.5$; it implies that the angular acceleration for $S = 0.3$ is comparably higher then $S = 0.5$ during the pitch-up cycle for the same reduced frequency. As discussed previously, the non-circulatory lift is the major component of the lift force, and it directly depends on the angular acceleration of the airfoil. Thus, the peak of the lift coefficient is massively high for the $S = 0.3$ during the pitch-up motion, then $S = 0.5$. The time history of the lift coefficient (Fig. 6(a)) for $A = 2°$ for distinct S values is similar in appearance with the angular acceleration (Fig. 1(b)), which is consistent with the previous discussion that the lift coefficient depends on the angular acceleration of the airfoil at a high reduced frequency. For the same parameters with $k = 9$, a similar trend is observed, although the magnitude of the peaks is considerably higher than $k = 3$.

Furthermore, Figures 6 (c) & 6(d) provides the force coefficients against the pitching angle of attack ($\alpha$) at Re = 10000 with $A = 2°$ and $k = 3$ and 9. The overall shape of the force coefficients vs. AoA plot for a given value of S is similar for all the parameters taken for the present study. As the airfoil motion changes from $S = 0.5$ to $S = 0.3$, the hysterics curve of force coefficients gets wider. The '*infinity*' shaped pattern for <$C_D$> formed for symmetric motion (S = 0.5) of the airfoil is distorted for the asymmetric (S = 0.3 and 0.4) sinusoidal motion.

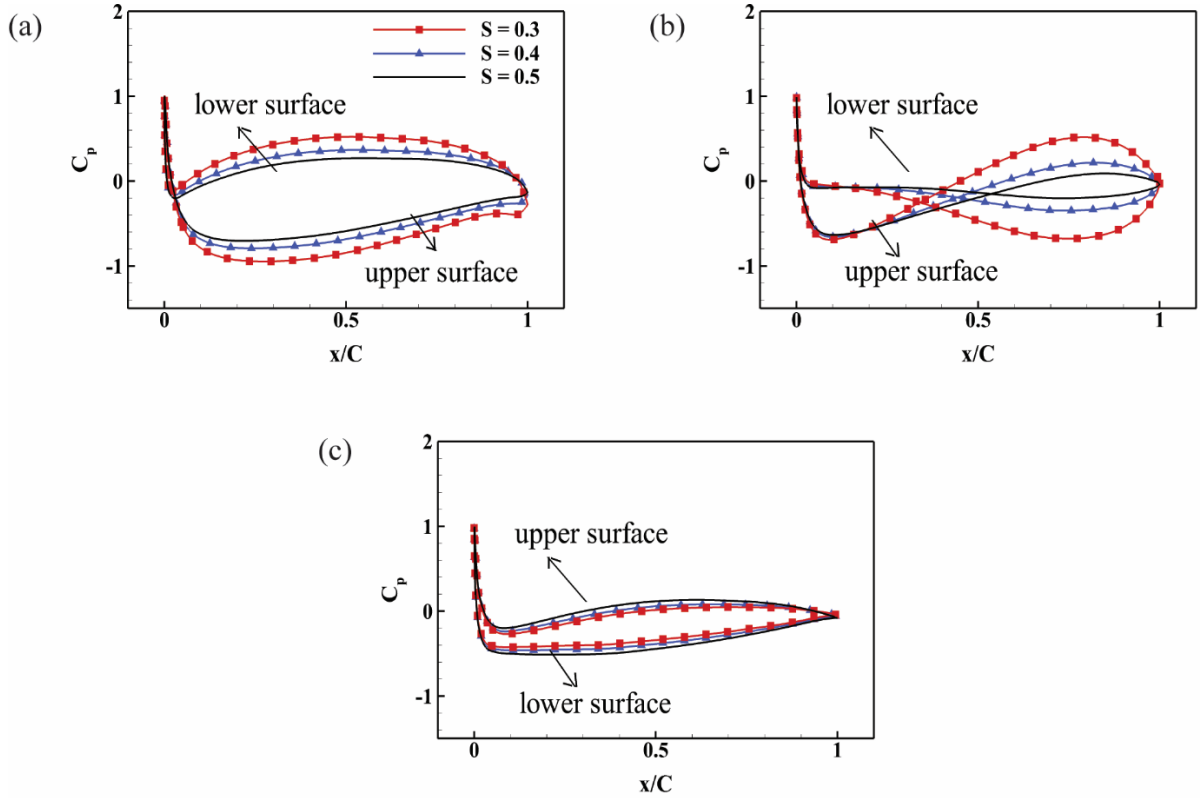

FIG. 7. The pressure coefficient (Cp) distribution over airfoil surface for A = 2°, $k$ = 3 and Re = 10000 for different S values. (a) 0° upstroke, (b) 1.73° upstroke and (c) 0° downstroke.

The pressure coefficient (Cp) over the airfoil surface is additionally investigated for the case A = 2°, Re = 10000, and $k$ = 3 for distinct S values. Fig. 7 reports the pressure coefficient for three different airfoil positions: (a) 0° upstroke, (b) 1.73° upstroke, and (c) 0° downstroke. Notably, the plot indicates the considerable effect of S on the pressure coefficient, which is useful to understand the instantaneous force coefficient behavior. As mentioned in Fig. 7(a), at 0° upstroke, the pressure coefficient on the airfoil upper side is negative, and for S = 0.3, it has minimum value as compared to S = 0.5. For the lower surface of the airfoil, the Cp value for S = 0.3 is maximum compared to S = 0.5. Due to the pressure distribution over the airfoil surface, at 0° upstroke, the instantaneous lift and drag coefficient for S = 0.3 has a higher value than S = 0.5, which can be seen from Fig. 6(c). The higher pitching rate during the pitching up motion for S = 0.3 is the reason for the higher surface pressure value. Similar pressure distribution is observed at 0° downstroke (Fig. 7(c)) besides the lower surface attains the negative pressure distribution that explains the negative instantaneous lift coefficient at that instance (Fig. 6(c)). The value of Cp for distinct S values is pretty much close, which also reflects from Fig. 6(c) that <$C_L$> is almost similar at 0° downstroke. Fig. 7(b) shows the pressure coefficient over the airfoil at 1.73° during the upstroke. The pressure distribution over the upper surface, near the leading edge, is negative and becoming almost zero towards the trailing edge for S = 0.5. Due to the intense negative pressure on the upper surface for S = 0.5, the lift coefficient is positive at 1.73° during the upstroke. Besides, for S = 0.3, the pressure coefficient attains the higher positive value on the upper surface, thereby resulting in negative instantaneous lift generation (Fig. 6(c)). Due to this, the instantaneous values of lift and drag coefficient are decreased at S = 0.3 than S = 0.5. Further, other parameters also exhibit similar pressure distribution, as reflected from the identical force coefficient.

### 4.4 Mean force coefficients

In continuation to the earlier discussion, one can notice that the force coefficients are periodic after the first cycle and approximately, the phase values of $C_L$ and $C_D$ are duplicated during each cycle; hence, the time-averaged mean force coefficients ($\overline{C_L}$ and $\overline{C_T}$) are calculated based on only the fourth cycle. The $\overline{C_L}$ and $\overline{C_T}$ (negative $\overline{C_D}$) for all the parameters considered in the present study are plotted against the reduced frequency $k$ in Figs. 8 and 9 for Re = 10000 and 3000, respectively. With the symmetric sinusoidal motion (S = 0.5), the mean thrust coefficient ($\overline{C_T}$) increases as $k$ increases for all the amplitudes (A = 2°, 4°, and 6°) at both Re = 10000 and 3000 (Figs. 8 and 9). The trend is similar to previous experimental observations for NACA0012 with the same order of parameters[13] but the magnitude of the $\overline{C_T}$ increases for SD7003 airfoil as compared to NACA0012. Simultaneously, the $\overline{C_L}$ declines as the reduced frequency vary from $k = 3$ to 9 for the symmetric pitching motion for all amplitudes and both Reynolds numbers. Although, for Re 3000 with amplitude A = 2°, the $\overline{C_L}$ remains almost constant for $k = 3$ and 6 and increases for $k = 9$ at S = 0.5.

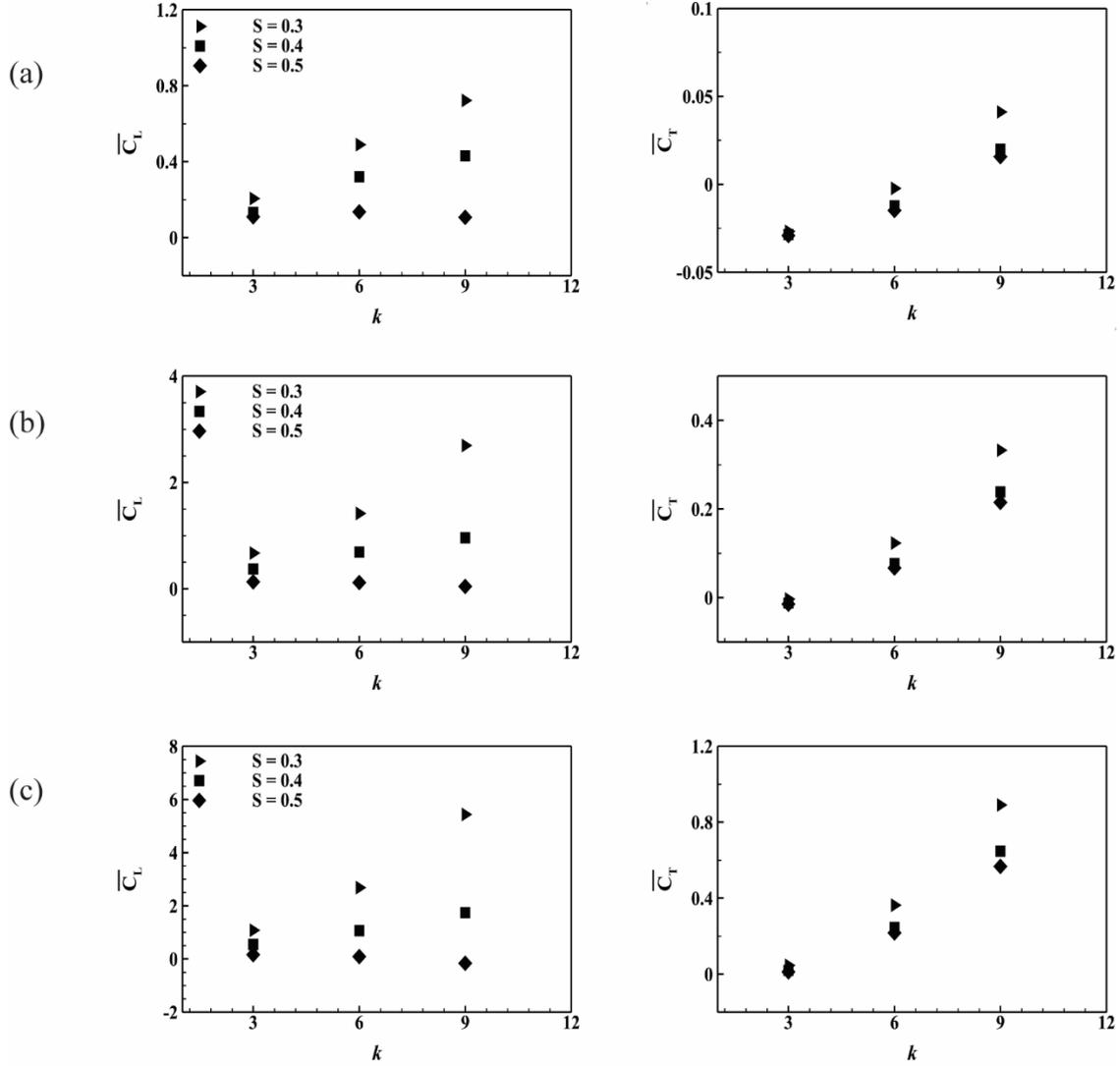

FIG. 8. The time-averaged mean force coefficient ($\overline{C_L}$ and $\overline{C_T}$) for Re = 10000 at different reduced frequency ($k$). (a) A = 2°, (b) A = 4° and (c) A = 6°.

The asymmetric sinusoidal motion (S = 0.3 and 0.4) has a remarkable effect on the $\overline{C_L}$ and $\overline{C_T}$ for both Re 10000 and 3000 (Figs. 8 and 9). The $\overline{C_T}$ raises significantly, two-three times for S = 0.3 as compared to S = 0.5 for all the amplitudes and the Reynolds numbers. An identical increment is observed in the $\overline{C_L}$ for the asymmetric pitching motion. The significant increment in $\overline{C_L}$ and $\overline{C_T}$ is due to the drastic increase in peaks of force coefficients as the airfoil goes under asymmetric sinusoidal motion (Fig. 6). The maximum $\overline{C_L}$ = 5.4385 and $\overline{C_T}$ = 0.8911 occur for the parameters combination with Reynolds number 10000, amplitude

A = 6°, reduced frequency $k = 9$ and asymmetric parameter S = 0.3. So, it may be concluded that the asymmetric sinusoidal motion enhances the mean force coefficients over the pitching airfoil for the same amplitude (A) and the reduced frequency ($k$).

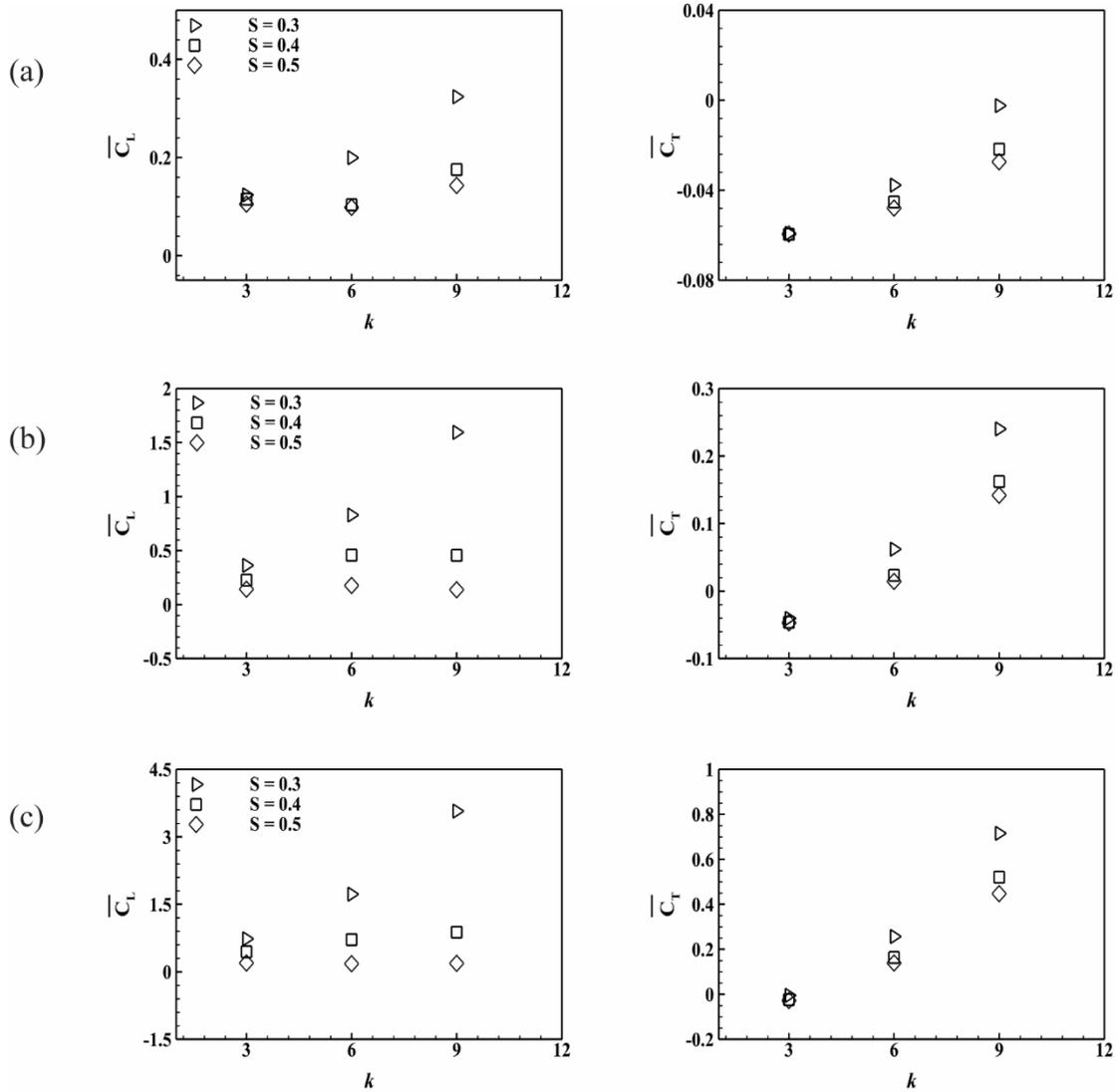

FIG. 9. The time-averaged mean force coefficient ($\overline{C_L}$ and $\overline{C_T}$) for Re = 3000 at different reduced frequencies ($k$). (a) A = 2°, (b) A = 4° and (c) A = 6°.

Above all discussions talk about the quantitative analysis which shows the impact of the unsteady parameters on the force coefficients. The effect of the amplitude and the reduced frequency on the mean thrust coefficient is similar to the NACA0012 for the same range of parameters, although the magnitudes are different[13, 14, 15]. Furthermore, the vorticity plot is shown in the next section to examine the influence of the unsteady parameters on the flow structures.

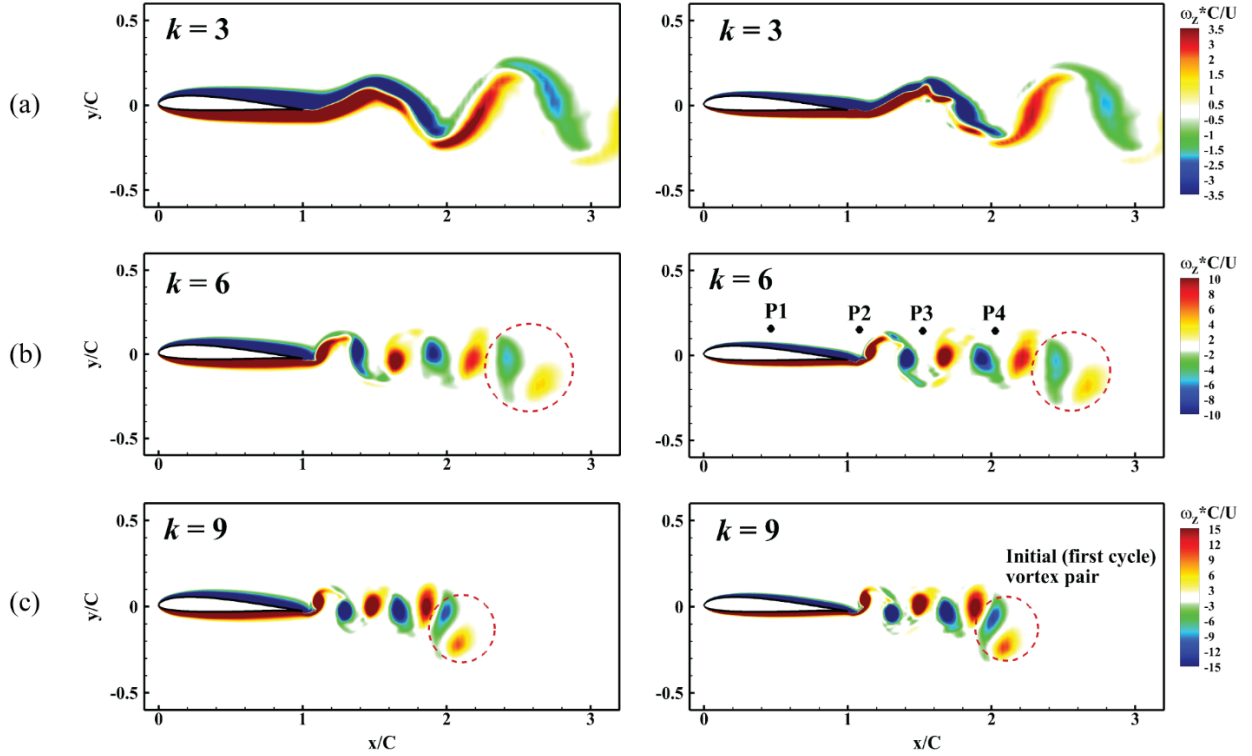

FIG. 10. The instantaneous normalized vorticity plot at the τ = 3.5 for A = 2° and symmetric sinusoidal motion (S = 0.5). Left column: Re 3000 and Right column: Re 10000. (a) $k = 3$, (b) $k = 6$ and (c) $k = 9$

### 4.5   Vorticity field

Figure 10 reports the instantaneous normalized vorticity field for symmetric (S = 0.5) sinusoidal motion at τ = 3.5 (at $A_{max}$) for both Reynolds number 3000 and 10000 with A = 2° for different $k$. The left column is for the Reynolds number 3000, whereas the right column shows the vorticity field for Re 10000. The clockwise vortex (CW, negative) and the counter-clockwise (CCW, positive) vortex pair detaches during each cycle. The vortex street pattern (reversed von Kármán) for both Reynolds numbers are similar for a given reduced frequency ($k$) which also indicates the nominal effect of the Reynolds number on the flow structure in the wake; albeit, the viscous boundary layer around the airfoil is thicker for Re = 3000 as compared to Re = 10000 due to the higher viscous effect at the low Reynolds number. As depicted in Fig.

10 (indicated by the dotted circle), the initial vortex pair (generated during the first cycle) is eccentric to the subsequent vortex pair, which elucidates the initial fluctuation in the flow field generating initial overshoot in the force coefficients measurement. The reason behind this initial fluctuation is the sudden acceleration of airfoil, as mentioned earlier.

Furthermore, the qualitative examination of the vorticity field reveals that as the reduced frequency increases, the alternative vortex pair becomes more compact in the downstream for both the Reynolds number (Fig. 10). For $k = 3$, there is only one vortex pair presented in the first chord (x/C = 2) downstream of the trailing edge while all the vortex pairs (generated up to $\tau = 3.5$) are closely packed in downstream at x/C = 2 for $k = 9$. Also, from the visualization of the contour of the vorticity, the strength of connecting a thin viscous layer (vortex braid) is reduced with increasing reduced frequency. These observations are consistent with the similar flow visualization results for NACA0012 at higher reduced frequencies ($k \geq 2$) of pitching airfoil for the Re 12000 by Bohl and Koochesfahani[15].

The asymmetric sinusoidal motion has drastically reshaped the vortex formation and vortex street pattern in the downstream. The instantaneous normalized vorticity is plotted during the time of one cycle (from $\tau = 3$ to $\tau = 4$) for Re 10000, amplitude A = 4° and $k = 6$ at different time instant for S = 0.5, 0.4 and 0.3. For better visualization, only the first chord (x/C = 2) distance in the downstream is taken into consideration (Fig. 11). The left column shows the vorticity field for S = 0.5, middle column for S = 0.4 and the right column for S = 0.3 at the same time instance. At the start of the fourth cycle (at $\tau = 3$), clockwise (CW) vortex from the previous cycle is attached to the trailing edge of the airfoil for all S values and the very next instant, as airfoil starts to pitch up, it detaches from the trailing edge and convects in the wake. At time $\tau = 3.1332$, the boundary layer around airfoil rolls and form a counter-clockwise (CCW) vortex at the trailing edge for all the S parameter's values.

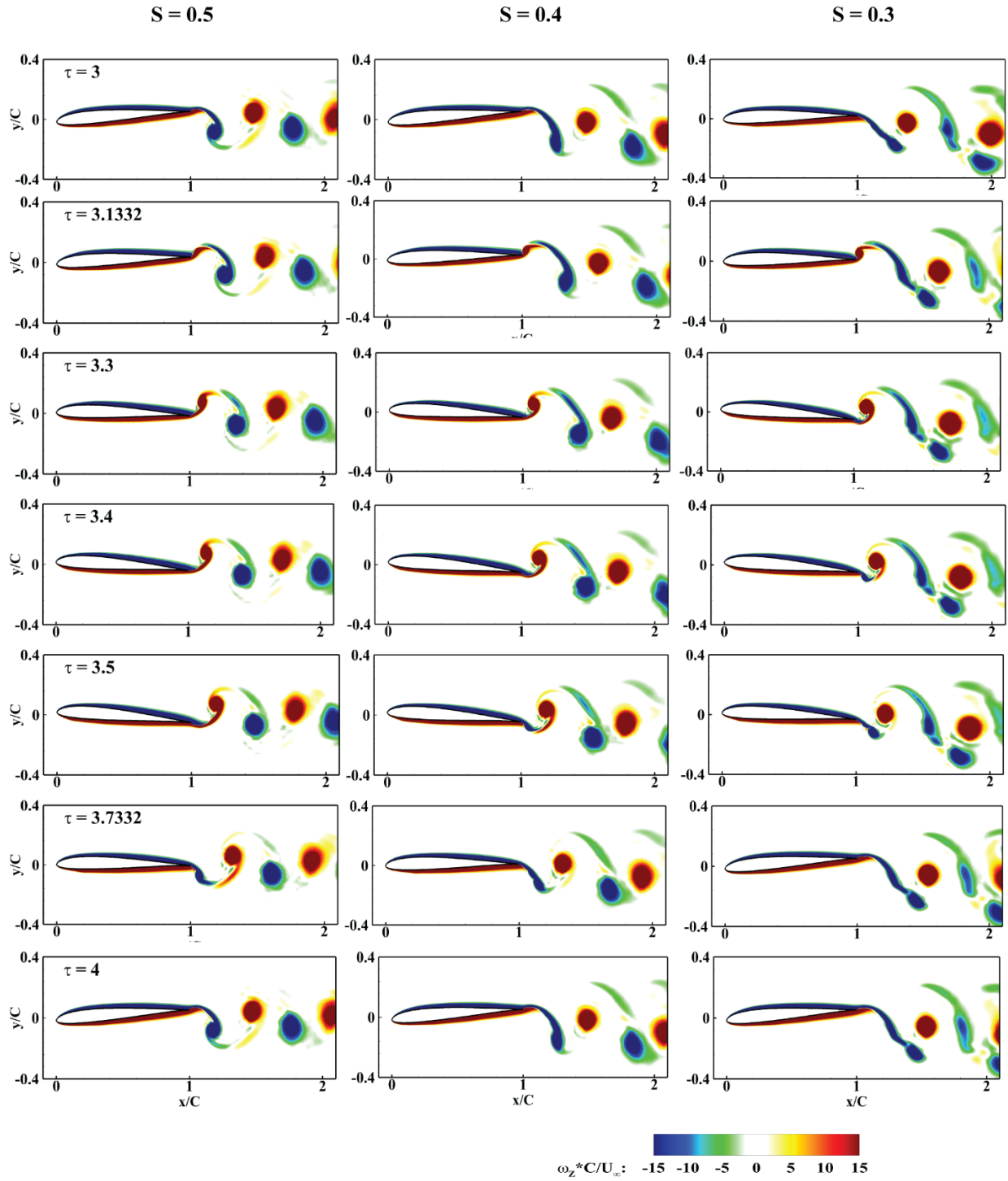

FIG. 11. The instantaneous normalized vorticity during the fourth cycle (from $\tau = 3$ to $\tau = 4$) at different time instance for A = 4°, $k = 6$, Re = 10000; left column: S = 0.5, Middle column: S = 0.4 and Right column: S = 0.3

Due to the higher acceleration during the pitch up cycle for S = 0.3, at $\tau$ = 3.3, the airfoil reaches $A_{max}$; the CCW vortex is fully grown and attached to the trailing edge. After reaching $A_{max}$, the airfoil starts moving down for S = 0.3, and the CCW vortex is about to detach the trailing edge, whereas for S = 0.4 and S = 0.5, the airfoil is still going under pitching up motion and CCW vortex is in the formation stage. The CCW vortex is fully developed at $\tau$ = 3.4 for S = 0.4 (airfoil is at $A_{max}$), and the airfoil is undergoing pitching up motion for S = 0.5. At the same time, the CCW vortex already detaches from the training edge and moves downstream for S = 0.3; simultaneously, the upper side boundary layer rolls and starts forming CW vortex at the trailing edge. At $\tau$ = 3.5, the airfoil is $A_{max}$ position for S = 0.5, and the CCW vortex is fully grown, but at the same instance, the CW vortex is continuously growing for asymmetric sinusoidal motion (S = 0.4 and 0.3). For the next instance at $\tau$ = 3.7332, the airfoil is undergoing pitch down motion, and the CW vortex is attached to the trailing edge for all the S values, but the CW vortex braids length and strength are different.

The comparisons of the vortex structure at $\tau$ = 4 with $\tau$ = 3 for all S values, it is qualitatively observed that the flow structures are similar for both the time instances, and it is established that the flow in the wake is periodic for each pitching cycle.

The unique vortex formation is observed for the asymmetry parameter S = 0.3. As we go from the symmetric sinusoidal motion (S = 0.5) to asymmetric sinusoidal motion (S = 0.3), the pitching down cycle becomes slower (Fig. 1(a)); as a result, it takes a longer time to complete the pitch-down motion. At $\tau$ = 3.4, the CCW vortex already detaches the trailing edge for S = 0.3, and the small CW vortex starts to form parallel to the CCW vortex at the trailing edge. Alongside, the viscous boundary layer on the upper side of the airfoil continuously rolls and increases the strength of the CW vortex. As the airfoil is pitching down extremely slow (for S = 0.3), the braid becomes strong enough to form another CW vortex and produces two CW vortices during the pitching down cycle. The primary CW vortex convects parallel with the CCW vortex in the wake though the secondary CW vortex convects in-line with the CCW vortex. As described in the above discussion about the negligible effect of the Reynolds number on the vortex structure formation and arrangement for the same parameters, a similar vortex formation over the pitching cycle is observed for Re 3000 with A = 4° and k = 6 too. The quantitative result (the time history of the force coefficients) and the vorticity filed indicate the periodic nature of the flow during each pitching cycle. This observation directs to look into the frequency of the flow parameters, which is carried out with the help of the Fast Fourier Transformation method.

### 4.6 Fast Fourier Transformation

The previous discussion supports the fact that the force coefficients and the flow structure are periodic for high reduced frequency ($k \geq 3$) and small amplitude (A ≤ 6°) pitching motion for LRN. The Fast Fourier Transformation (FFT) is an efficient tool to investigate the frequency of the periodic flow. The Fast Fourier Transformation is performed for the case A = 4° and Re 10000; for $k$ = 3, 6, and 9 with symmetric sinusoidal motion, and for $k = 6$ with asymmetric sinusoidal motion. As shown in Fig. 2(b), four probes are located (also see Fig. 10(b)); one is above the airfoil at mid-span (P1), which is not affected by the viscous boundary layer and the other three (P2, P3, and P4) are downstream of the trailing edge. The pressure values are recorded for a time of 20 cycles which is sufficient to perform FFT. At τ = 3.5, the initial vortex pair almost passes through one chord distance downstream of the trailing edge (Fig. 10) for all the reduced frequency. Therefore, the FFT is calculated from the fifth cycle (τ = 5) after the flow is entirely periodic at all the probe locations.

TABLE VI. The reduced frequency ($k$) with the calculated FFT frequency and reduced frequency ($k^*$) at P1 for A = 4° and Re = 10000 with symmetric sinusoidal motion (S = 0.5).

| Pitching reduced frequency ($k$) | FFT frequency $f$ (Hz) | FFT reduced frequency ($k^*$) |
|---|---|---|
| 3 | 0.145 | 3.03 |
| 6 | 0.2899 | 6.07 |
| 9 | 0.4272 | 8.94 |

Figure 12 shows the FFT of the pressure signal for various cases for different *k* with symmetric sinusoidal motion (S = 0.5). At probe location P1, there is only one frequency for all *k*, so the variation of the pressure at P1 is entirely harmonic. At the downstream location (P2, P3, and P4), there are other one or two frequencies, which are the second and third harmonic of the most dominating frequency present as a consequence of vortex shedding. Although, for *k* = 3, the second harmonic frequency is dominant over the other frequencies in the downstream.

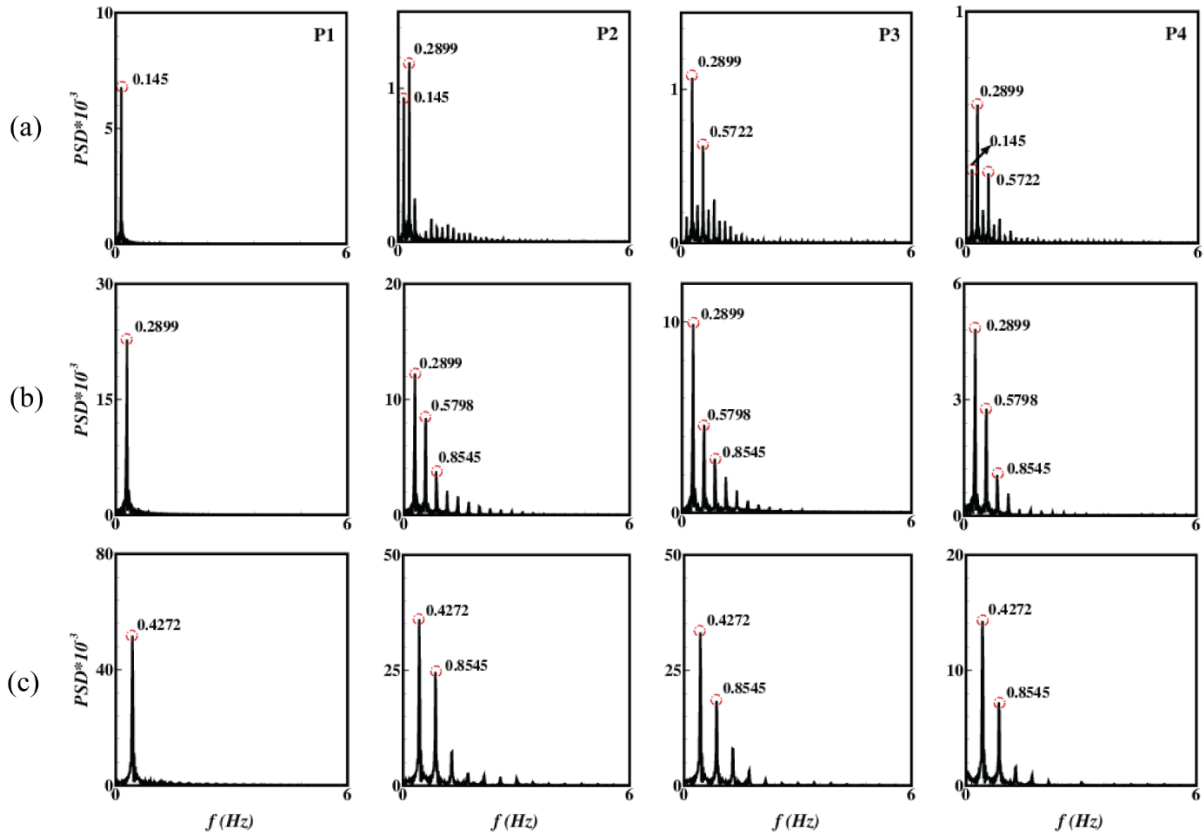

FIG. 12. The Fast Fourier Transformation of pressure signal at the probe location P1, P2, P3 and P4 for Re = 10000, A = 4°, S = 0.5 (symmetric sinusoidal motion). (a) *k* = 3, (b) *k* = 6 and (c) *k* = 9.

Based on the inlet velocity $U_\infty$, the FFT frequency at the P1 location is converted into the FFT reduced frequency ($k^*$), which is defined as $k^* = 2\pi f C / 2U_\infty$, where *f* is the FFT frequency, and C is the chord length of the airfoil. The FFT frequencies are 0.145, 0.2899, and 0.4272 Hz for the pitching reduced frequency (*k*) 3, 6, and 9, respectively (Fig. 12) at the location P1. The $k^*$ based on the FFT and corresponding reduced frequencies ($k$) of the pitching airfoil are listed in Table VI. It is seen that for all reduced frequency cases, at location P1, the FFT reduced frequency ($k^*$) is approximately the same with the pitching reduced frequency ($k$). These observations have contributed considerably towards the conclusion that in the flow domain which is not affected by the viscous boundary layer or any vortex structure, the flow property is governed by the airfoil pitching frequency. As a result, the lift coefficient for a higher reduced frequency pitching airfoil is dominated by the frequency (angular acceleration ~ $k^2$) of the airfoil's pitching motion.

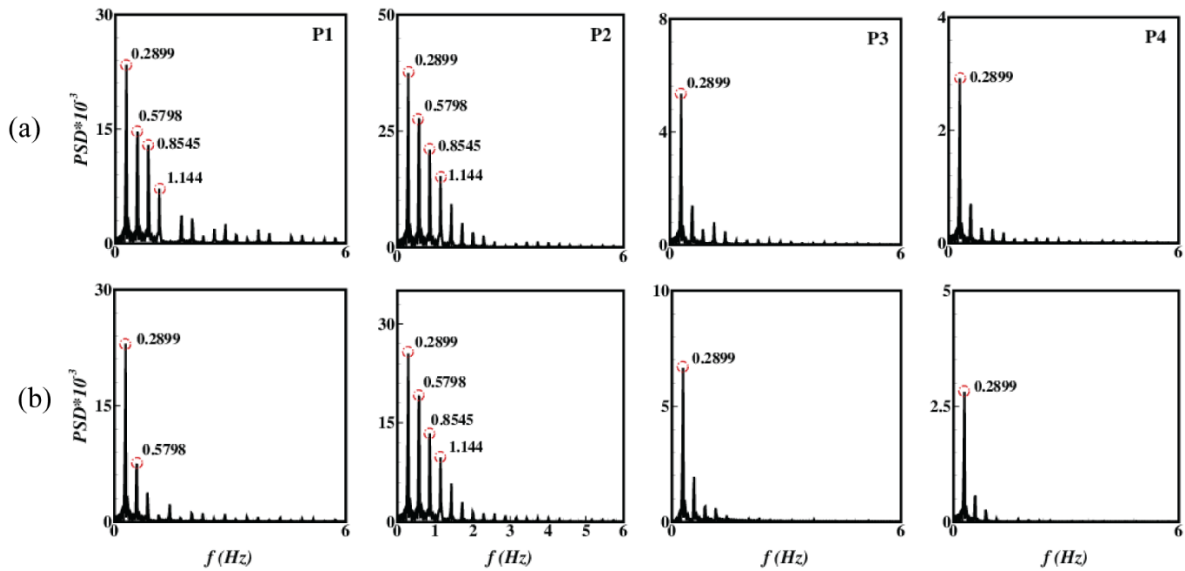

FIG. 13. The Fast Fourier Transformation of pressure signal at the location P1, P2, P3 and P4 for Re 10000, A = 4°, $k$ = 6 for asymmetric sinusoidal motion. (a) S = 0.3 (b) S = 0.4.

For the asymmetric sinusoidal motion (S = 0.3 and 0.4), Figure 13 reports the FFT for the same probe locations, as mentioned earlier with $k$ = 6. It is observed that at P1 location for S = 0.3 (Fig. 13(a)), there is more than one dominant frequency, i.e., second, third, and fourth harmonic frequencies, present due to the asymmetric pitching motion of the airfoil. Whereas for the asymmetric motion S = 0.4, only second harmonic frequency has a significant impact at P1 (Fig. 13(b)). However, for both cases, i.e. S = 0.3 and S = 0.4, the dominating frequency is 0.2899 Hz ($k^*$ = 6.07), and as mentioned above, it is equivalent to the airfoil's pitching reduced frequency ($k$ = 6). As moving to the downstream, near the airfoil trailing edge (at P2), the similar frequencies are observed for both S values 0.3 and 0.4, which shows that near the airfoil, the flow dynamics are dominated by the airfoil motion at a higher reduced frequency. In the wake, at the location P3 and P4, there is only one dominant frequency present $f = 0.2899$ Hz ($k^* = 6.07$), and the higher harmonic frequencies are insignificant.

5. **Conclusion**

The numerical calculations are carried out over low Reynolds number airfoil SD7003 for different sets of parameters (A = 2°, 4°, 6°; $k$ = 3, 6, 9, and S = 0.3, 0,4, 0.5) for two Reynolds number 3000 and 10000. Additionally, a theoretical model is invoked to calculate the lift coefficient and to compare it with the numerical result for the symmetric pitching motion. The influence of the lift components, non-circulatory lift, and circulatory lift components has been reported through theoretical analysis. The theoretical analysis shows that at a higher reduced frequency, the lift generation is majorly due to the non-circulatory lift which depends on the angular acceleration ($\sim k^2$) of the airfoil. This implies that the inertial effect dominates over the viscous effect on the lift generation at high reduced pitching frequency and a low Reynolds number flow. The theoretical lift coefficient is in excellent agreement with the numerical result at the low amplitude (A=2°). Hence, the inviscid theory can still be used to predict the lift coefficient accurately for low amplitude, high reduced frequency pitching motion at LRN. However, the computational or experimental studies are required for high amplitude (A=6°) to calculate the lift coefficient.

Moreover, the lift coefficient peaks are approximately independent of the Reynolds number irrespective of asymmetry in a pitching motion. The results show that the asymmetry in the pitching motion affects the force coefficients, the pressure coefficient, and mean lift, as well as the thrust coefficient for the same A and $k$. Furthermore, the flow structures for different Re are almost similar in the wake for a fixed A and $k$, which indicates a negligible effect of Re on the flow structure in the downstream. It is also observed that the asymmetric sinusoidal motion has a considerable influence on the formation of the vortex pair and strength of the braid. The FFT analysis reveals that for the symmetric as well as asymmetric sinusoidal motion of the airfoil, the dominant frequency in the flow is the same as the airfoil pitching reduced frequency ($k^* = k$).

## Acknowledgment

Simulations are carried out on the computers provided by the Indian Institute of Technology Kanpur (IITK) (www.iitk.ac.in/cc), and the manuscript preparation, as well as data analysis, has been carried out using the resources available at IITK. The support is gratefully acknowledged.Simulations are carried out on the computers provided by the Indian Institute of Technology Kanpur (IITK) (www.iitk.ac.in/cc), and the manuscript preparation, as well as data analysis, has been carried out using the resources available at IITK. The support is gratefully acknowledged.

## Data Availability Statement

The data that support the findings of this study are available from the corresponding author upon reasonable request.